\documentclass[aps,prb,twocolumn]{revtex4}
\usepackage{amsfonts}
\usepackage{amsmath}
\usepackage{amssymb}
\usepackage{graphicx}
\begin{document}
\title{Periodic alternating $0,\pi$-junction structures as realization of $\varphi$-Josephson junctions}
\author{A. Buzdin}
\affiliation{CPMOH, UMR 5798, Universit\'{e} Bordeaux 1, 33405
Talence Cedex, France}
\author{A.\ E.\ Koshelev}
\affiliation{Materials Science Division, Argonne National Laboratory, Argonne, Illinois 60439}

\begin{abstract}
We consider the properties of a periodic structure consisting of
small alternating $0$- and $\pi$- Josephson junctions. We show
that depending on the relation between the lengths of the
individual junctions, this system can be either in the homogeneous
or in the phase-modulated state. The modulated phase appears via a
second order phase transition when the mismatch between the
lengths of the individual junctions exceeds the critical value.
The screening length diverges at the transition point. In the
modulated state, the equilibrium phase difference in the structure
can take any value from $-\pi$ to $\pi$ ($\varphi$-junction).
The current-phase relation in this structure has very unusual
shape with two maxima. As a consequence, the field dependence of
the critical current in a small structure is very different from
the standard Fraunhofer dependence. The Josephson vortex in a long
structure carries partial magnetic flux, which is determined by
the equilibrium phase.

\end{abstract}
\maketitle

The minimum energy of Josephson junction (JJ) usually corresponds
to the zero phase difference of the superconducting order
parameter \cite{JJTextbook}. However, long time ago it has been
predicted that the JJs with ferromagnetic interlayer (S-F-S
junctions) may have ground state with phase difference equal to
$\pi$ ("$\pi$-junctions")\cite{SFSpi-junction}. Only recently this
prediction has been experimentally verified
\cite{ryazanov,Kontos2,Blum} and now the controllable fabrication
of S-F-S $\pi$-junctions becomes possible. Another realization of
$\pi$-junctions became possible due to the d-wave symmetry of the
order parameter in high-temperature superconductors
\cite{Kirtley,VanHarlingen}. Recent experiments on
YBa$_{2}$Cu$_{3}$0$_{7}$-Nb ramp long JJs of zigzag geometry
demonstrated that such junctions are composed of alternating
facets of $0$- and $\pi$- junctions\cite{Smilde,Mints}. In a
pioneering work \cite{Kuzii} Bulaevskii \emph{et al.} demonstrated
that the spontaneous Josephson vortex (JV) carrying flux
$\Phi_{0}/2$ appears at the boundary between $0$- and $\pi$- JJs.
The structure of such semifluxon has been studied in more detail
in Refs.\ \onlinecite{Xu,Goldobin}.

In the present work we study the properties of the periodic array
of $0$- and $\pi$- JJs. We show that depending on the ratio of
lengths of $0$- and $\pi$- JJs, such an array can be either in the
homogeneous or modulated state. The second order phase transition
between these states takes place when the length mismatch between
$0$- and $\pi$- JJs is small. At the transition point the
screening length of the magnetic field diverges.
In the modulated states the average phase difference $\varphi_{0}$
can take any value between $-\pi$ and $\pi$. Further, we call such
systems $\varphi $-junctions. Such structures were first predicted
by Mints \cite{MintsPRB98} in the case of randomly alternating
$0$- and $\pi$- JJs. We study the properties of
$\varphi$-junctions: effective Josephson current, Josephson
length, and  Fraunhofer-like oscillations of the critical current.
Magnetic properties of long $\varphi$-junctions are determined by
two types of Josephson vortices carrying partial fluxes
$\Phi_{0}\varphi_{0}/\pi$ and $\Phi_{0}(\pi -\varphi_{0})/\pi$
(see also \cite{MintsPRB98,MintsPRB01}). We find analytically the
shapes of these unusual Josephson vortices. We also demonstrate
that at the boundary between $\varphi$- and usual JJs a Josephson
vortex appears carrying a partial flux $\Phi
_{0}\varphi_{0}/2\pi$.

\begin{figure}
\includegraphics[width=3.4in,clip]{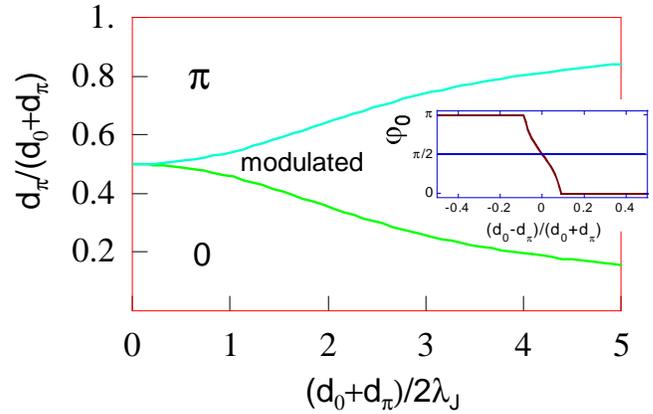}
\caption{Phase diagram of the periodic alternating
$0$-$\pi$ junction structure. Inset shows dependence of the
equilibrium phase on the length mismatch for
$d_{0}d_{\pi}=\lambda_J^2$} \label{Fig-PhaseDiag}
\end{figure}
Consider a periodic structure composed of alternating $0$- and
$\pi$-JJs of lengths $d_{0}$ and $d_{\pi}$ respectively. The
energy per period of such a structure is given by
\begin{align}
F &  =\frac{\Phi_{0}j_{c}}{2\pi c}\int_{-d_{0}}^{0}\left(  \frac{\lambda
_{J}^{2}}{2}\left(  \frac{d\phi}{dx}\right)  ^{2}+1-\cos\phi\right)
dx\nonumber\\
&  +\frac{\Phi_{0}j_{c}}{2\pi c}\int_{0}^{d_{\pi}}\left(  \frac{\lambda
_{J}^{2}}{2}\left(  \frac{d\phi}{dx}\right)  ^{2}+1+\cos\phi\right)
dx,\label{Energy}
\end{align}
where $j_{c}$ is the Josephson current density, $\lambda_{J}$ is
the Josephson length of the individual junctions,
$\lambda_{J}^{2}=c\Phi_{0}/(8\pi^{2} tj_{c})$, and
$t=t_{0}+\lambda_{1}+\lambda_{2}$ is the effective junction
thickness \cite{JJTextbook}. For simplicity, we focus on the case
when the critical current densities $j_{c}$ are the same in $0$-
and $\pi$-JJs. Generalization for different critical currents is
straightforward. Possible realization of such system is the zigzag
junction between high-T$_{c}$ and conventional superconductors as
well as S-F-S junctions with periodically modulated thickness of
the ferromagnetic interlayer. In the case of the zigzag structure,
$x$ is the coordinate along the zigzag boundary. The ground state
phase distribution is determined by the equation
\begin{equation}
\frac{d^{2}\phi}{dx^{2}}+j(x)\sin\phi =0,\
-d_{0}<x<d_{\pi},\label{PhaseEquation}
\end{equation}
where $j(x)=-\lambda_{J}^{-2}$ in $0$-JJs and
$j(x)=\lambda_{J}^{-2}$ in $\pi$-JJs.  At the boundaries, the
phase and its derivative have to be continuous. Periodicity of
$\phi(x)$ implies that the solution is symmetric with respect to
$x=-d_{0}/2$ and $d_{\pi}/2$, i.e.,
$d\phi/dx\vert_{x=-d_{0}/2}=d\phi /dx\vert_{x=d_{\pi}/2}=0$. Eq.\
(\ref{PhaseEquation}) always has homogeneous solutions $\phi=0$
and $\phi=\pi$. However these solutions give the ground state only
in some range of the ratio $d_{0}/d_{\pi}$. In general, Eq.
(\ref{PhaseEquation}) also allows for the inhomogeneous solution.
Consider the case $d_{\pi}<d_{0}$ and weak modulation around the
$\phi=0$ state. In this case Eq.\ (\ref{PhaseEquation}) can be
linearized and its solution is given by
\begin{equation}
\phi= \genfrac{\{}{.}{0pt}{}{A\cosh\frac{x+d_{0}/2}{\lambda_{J}},\
-d_{0} <x<0,}{B\cos\frac{x-d_{\pi}/2}{\lambda_{J}},\ 0<x<d_{\pi}.}
\end{equation}
Matching the logarithmic derivative $d\ln\phi/dx$ of these
solutions at $x=0$, we obtain the following condition
\begin{equation}
\tanh(d_{0}/2\lambda_{J})=\tan(d_{\pi}/2\lambda_{J})
,\label{Condition}
\end{equation}
for the onset of the modulated solution. The energy analysis shows
that the uniform $\phi=0$ solution is favorable at $\tanh\left(
d_{0}/2\lambda_{J}\right)  >\tan\left( d_{\pi}/2\lambda
_{J}\right)  $.  In the opposite case, $d_{\pi}>d_{0}$, the
uniform $\phi=\pi$ solution is favorable at $\tanh\left(  d_{\pi
}/2\lambda_{J}\right)
>\tan\left(  d_{0}/2\lambda_{J}\right)$. In the case of a finite
junction split into $0$- and $\pi$- pieces, condition
(\ref{Condition}) was first derived in Ref.\ \onlinecite{Kuzii}.
These results are summarized in the phase diagram shown in Fig.
\ref{Fig-PhaseDiag}. In the limit $d_{0}\gg\lambda_{J}$ the
condition for the modulated solution is given by
$d_{\pi}>(\pi/2)\lambda_{J}$. In the case
$d_{0},d_{\pi}\ll\lambda_{J}$ the region of the modulated solution
is given by
\begin{equation}
\frac{\left\vert d_{0}-d_{\pi}\right\vert }{d_{0}}<\frac{d_{0}^{2}}
{6\lambda_{J}^{2}}\ll1,\label{ModCondition}
\end{equation}
i.e., the modulated solution exists only for a very small length
mismatch. Since $\lambda_{J}$ is temperature dependent, the
transition into the modulated state may occur with decreasing
temperature.
%

\begin{figure}
\includegraphics[width=3.4in,clip]{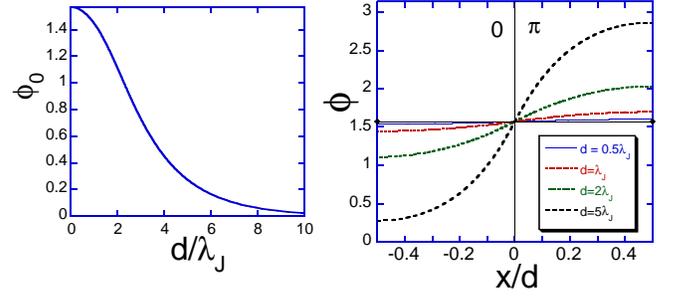}
\caption{\emph{Left panel}: dependence of the minimum phase on the
length of junction for $d_0=d_{\pi}$. \emph{Right panel}: shape of
phase modulation at different junctions sizes} \label{Fig-phi_x}
\end{figure}

In the case $d_{0}=d_{\pi}\equiv d$, the phase distribution is
always modulated. Lets find this distribution. From symmetry
$\phi(-x)=\pi-\phi(x)$, in particular, $\phi(0)=\pi/2$. The first
integral of Eq.\ (\ref{PhaseEquation} ) for $x<0$ is given by
\begin{equation}
\frac{\lambda_{J}^{2}}{2}\left(  \frac{d\phi}{dx}\right)  ^{2}+\cos\phi
=\cos\phi_{0}
\end{equation}
with $\phi_{0}=\phi(-d/2)$. The solution is given by
\begin{equation}
x(\phi)=-\frac{\lambda_{J}}{\sqrt{2}}\int_{\phi}^{\pi/2}\frac{d\phi^{\prime}
}{\sqrt{\cos\phi_{0}-\cos\phi^{\prime}}},
\end{equation}
where $\phi_{0}$ is determined by the condition
$x(\phi_{0})=-d/2$. Dependence $\phi_{0}(d/\lambda_{J})$ and
shapes of the phase variation $\phi(x)$ for different values of
$d/\lambda_{J}$ are presented in Fig.\ \ref{Fig-phi_x}. For small
length $d\ll\lambda_{J}$, it is easy to find that the phase
distribution \ for $-d<x<0$ is $\phi(x)=\pi/2+x(x+d)/\left(
8\lambda_{J} ^{2}\right)  $, i.e., it varies weakly around
$\pi/2$.
%

Consider a structure composed of short alternating $0$ and $\pi$-
JJs with a small length mismatch, $|d_{0}-d_{\pi}|\ll d_{0}$. In
this case the function $j(x)$ in Eq.\ (\ref{PhaseEquation})
oscillates rapidly and
we can present $\phi(x)$ as $\phi(x)=\varphi(x)+\xi(x)$, where the
function $\varphi(x)$ varies slowly over the distance $d_{0}$ and
$\xi(x)$  oscillates rapidly with $\overline{\xi (x)}=0$. In the
limit $d_{0,\pi}< \lambda_J$, the oscillating term is small (as
one can see from the right panel of Fig.\ \ref{Fig-phi_x} even at
$d_{0},d_{\pi}= \lambda_J$, (i)$|\xi(x)|\leq 0.13\ll 1$ and (ii)
the oscillating part is smaller than the average phase). In this
case one can use a coarse-grained approximation (see, e.g., Ref.
\onlinecite{LanLifMech}),  meaning that we can average out the
rapidly oscillating function $\xi(x)$ and derive the equation for
the slowly varying function $\varphi(x)$ only:
\begin{equation}
\frac{d^{2}\varphi}{dx^{2}}=-\frac{d_{0}d_{\pi}}{24\lambda_{J}^{4}}\sin\left(
2\varphi\right)
+\frac{d_{0}-d_{\pi}}{(d_{0}+d_{\pi})\lambda_{J}^{2}}\sin\left(
\varphi\right)  ,\label{PhaseEq}
\end{equation}
Here the first term in the right hand side appears due to current
modulations and the second term is proportional to the average
Josephson current. A similar equation was derived in Ref.\
\cite{MintsPRB98}. The neglected terms in Eq.\ (\ref{PhaseEq}) are
smaller by a factor of $(d_{0}/\lambda_J)^2$. Without the external
magnetic field, when condition (\ref{ModCondition}) is fulfilled,
the equilibrium phase difference $\varphi_{0}$ is:
\begin{equation}
\cos\varphi_{0}=f\equiv\frac{12\lambda_{J}^{2}\left(
d_{0}-d_{\pi}\right)}{d_{0}d_{\pi}(d_{0}+d_{\pi})},
\label{Equil_phi_0}
\end{equation}
and for $|f|<1$, the average phase difference lies in the region
$0<\left\vert \varphi_{0}\right\vert <\pi$,i.\ e.\, \emph{we have
the realization of a JJ with arbitrary ground state phase
difference (}$\varphi $\emph{-junction)}. For $|f|>1$ homogeneous
$0$- or $\pi$- phase is realized. An example of the dependence of
the equilibrium phase $\varphi_{0}$ on length mismatch is shown in
the inset of Fig.\ \ref{Fig-PhaseDiag}.
\begin{figure}
\includegraphics[width=3.4in,clip]{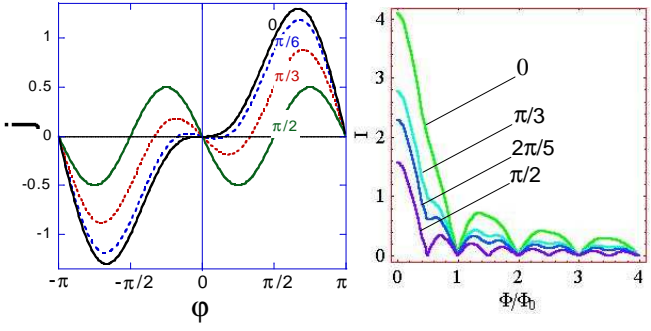}
\caption{\emph{Left panel}: current-phase relationships for
different values of the equilibrium phase $\varphi_{0}$ (see Eq.\
(\ref{Phase-Current})). Curves are marked by the value of
$\varphi_{0}$. Unit of current density $j$ in this plot is
$d_{0}^{2}j_{c}/(12\lambda_{J} ^{2})$. \emph{Right panel}:
dependencies of critical current of small $0,\pi$- structure on
the magnetic flux $\Phi$ through the structure, for different
values of $\varphi_{0}$. The current unit at the vertical axis is
$j_{J}Sd_{0}^{2}/(12\lambda_{J}^{2})$, where $S$ is the cross
section area of the structure. } \label{Fig-j_phi_j_h}
\end{figure}

The energy in terms of the coarse-grained phase is given by
\begin{equation}
F=\frac{\Phi_{0}j_{c}\lambda_{J}^{2}}{2\pi c}\int\left(
\frac{1}{2}\left( \frac{d\varphi}{dx}\right)  ^{2}+\frac{\left(
\cos\varphi-f\right)  ^{2} }{2\lambda_{\varphi}^{2}}\right)  dx,
\end{equation}
with
$\lambda_{\varphi}^{2}=12\lambda_{J}^{4}/d_{0}^{2}\gg\lambda_{J}^{2}$.
The coarse-grained Josephson current density flowing through the
array is given by
\begin{equation}
j(\varphi)=\frac{d_{0}^{2}j_{c}}{12\lambda_{J}^{2}}\sin\left(  \varphi\right)
\left(  f-\cos\varphi\right)  .\label{Phase-Current}
\end{equation}
This current-phase relation is quite peculiar (see left panel in
Fig.\ \ref{Fig-j_phi_j_h}). The current has two maxima and two
minima, which are achieved at $\cos\varphi_{\pm}=\left(
f\pm\sqrt{f^{2}+8}\right)  /4$. The absolute maximum
\[
j_{\max}=\frac{d_{0}^{2}j_{c}}{48\lambda_{J}^{2}}\sqrt{\frac{1}{2}-\frac{f}
{8}\left(  f-\sqrt{f^{2}+8}\right)  }\left(
3f+\sqrt{f^{2}+8}\right)
\]
is achieved at $\varphi=\varphi_{-} $.

Consider screening of the weak magnetic field by the periodic
structure. In the modulated phase linearizing Eq.\ (\ref{PhaseEq})
near $\varphi_{0}$, $\varphi=\varphi_{0}+\tilde{\varphi}$, we
obtain a linear equation for the phase variation $\tilde{\varphi}$
\begin{equation}
\frac{d^{2}\tilde{\varphi}}{dx^{2}}=\frac{\sin^{2}\varphi_{0}}{\lambda
_{\varphi}^{2}}\tilde{\varphi},
\end{equation}
One can see that the effective screening length
\begin{equation}
\lambda_{\mathrm{eff}}\equiv\lambda_{\varphi}/\sin\varphi_{0}=\lambda
_{\varphi}/\sqrt{1-f^{2}}\text{, for }|f|<1
\end{equation}
diverges at the transition point. In the homogeneous phase, $|f|>1$, similar
derivation gives
\begin{equation}
\lambda_{\mathrm{eff}}=\lambda_{\varphi}/\sqrt{|f|-1}\text{, for
}|f|>1.
\end{equation}
Because $\lambda_{\varphi}(T)$ diverges at $T\rightarrow T_{c}$
faster than $f(T)$, $\lambda _{\mathrm{eff}}(T)$ is nonmonotonic
above the transition point. The minimum value of
$\lambda_{\mathrm{eff}}$, $\lambda_{\mathrm{eff}}=2d_{0}^{2}/\sqrt
{3}\left( d_{0}-d_{\pi}\right)  $ is reached at $f=2$

Exactly at the transition point, $f=1$, screening is nonlinear and
the phase variation obeys equation
\begin{equation}
2\lambda_{\varphi}^{2}d^{2}\tilde{\varphi}/dx^{2}=\tilde{\varphi}^{3}.
\end{equation}
The general solution of this equation is given by
$\tilde{\varphi}=\pm 2\lambda_{\varphi}/\left(  x+C\right)  $.
Using the phase-field relation $H=(\Phi_{0}/2\pi
t)d\tilde{\varphi}/dx$, we derive from this solution that the
magnetic field decays inside the structure as
\begin{equation}
H(x)=H_{0}\left(  1+x/\sqrt{\frac{\lambda_{\varphi}\Phi_{0}}{\pi tH_{0}}
}\right)  ^{-2}
\end{equation}
where $H_{0}$ is the external magnetic field. It is interesting to
note that the field tail at large $x$ does not depend on the
applied field
\begin{equation}
H(x)=\frac{\lambda_{\varphi}\Phi_{0}}{\pi tx^{2}}\text{, for
}x\gg\sqrt {\frac{\lambda_{\varphi}\Phi_{0}}{\pi tH_{0}}}.
\end{equation}

In addition to the usual $2\pi$-degeneracy, in the case $f<1$, the
ground state energy is also degenerate with respect to sign change
of the equilibrium phase $\varphi_{0}\rightarrow-\varphi_{0}$.
This degeneracy leads to the appearance of two new kinds of
solitons in which the phase sweeps either from $-\varphi_{0}$ to
$\varphi_{0}$ or from $\varphi_{0}$ to $2\pi-\varphi_{0}$.
Rewriting  Eq.\ (\ref{PhaseEq}) in a more convenient form
\[
\frac{d^{2}\varphi}{dx^{2}}=\frac{1}{\lambda_{\varphi}^{2}}\sin\left(
\varphi\right)  \left(  \cos\varphi_{0}-\cos\varphi\right),
\]
we derive the shapes of the two solitons
\begin{subequations}
\begin{align}
\varphi_{1}(x)  &  =2\arctan\left(  \tanh\left(
\frac{x\sin\varphi_{0} }{2\lambda_{\varphi}}\right)  \tan\left(
\frac{\varphi_{0}}{2}\right)
\right)  ,\label{Shape1}\\
\varphi_{2}(x)  &  =\pi+2\arctan\left(  \frac{\tanh\left(  x\sin\varphi
_{0}/2\lambda_{\varphi}\right)  }{\tan\left(  \varphi_{0}/2\right)  }\right)
. \label{Shape2}
\end{align}
The soliton energies per unit length are given by
\end{subequations}
\begin{subequations}
\begin{align}
\varepsilon_{1}  & =\varepsilon_{J0}(\lambda_{J}/4\lambda_{\phi}
)\left( \left\vert \sin\varphi_{0}\right\vert
-\varphi_{0}\cos\varphi
_{0}\right)  ,\ \label{En_Sol}\\
\varepsilon_{2}  & =\varepsilon_{J0}(\lambda_{J}/4\lambda_{\phi})
\left( \left\vert \sin\varphi_{0}\right\vert +\left(  \pi-\varphi
_{0}\right)  \cos\varphi_{0}\right),
\end{align}
where $\varepsilon_{J0}=4\Phi_{0}j_{c}\lambda_{J}/(\pi c)$ is the
energy of a single soliton in a uniform JJ. At the transition
point, $f=1$, the first soliton vanishes and the second one
acquires the following shape, $\varphi _{2}(x)=\pi+2\arctan\left(
x/\lambda_{\varphi}\right) $. Even though for $d_{0}>d_{\pi}$ the
first energy is smaller than the second one, both solitons play a
role in the magnetic properties of the $\varphi$-junction. This is
because without the second solitons the system can not contain two
or more first solitons of the same sign due to topological
constrains. In the external magnetic field $H$, the energy of the
system with a small number of solitons of both kinds, $N_1$ and
$N_2$, can be written as
\end{subequations}
\begin{equation}
E=\varepsilon_{1}N_{1}+\varepsilon_{2}N_{2}-\left(
N_{1}\frac{\varphi_{0}
}{\pi}+N_{2}\frac{\pi-\varphi_{0}}{\pi}\right)
\frac{H\Phi_{0}}{4\pi},
\end{equation}
and has to be supplemented by the topological constrain
$\left\vert N_{1}-N_{2}\right\vert \leq1$. The penetration
scenario is determined by the ratio of the soliton energy
$\varepsilon_{\alpha}$ to its flux $\Phi_{\alpha} $. Simple
analysis shows that for $d_{0}>d_{\pi}$ these ratios are arranged
in the order
$\varepsilon_{1}/\Phi_{1}<(\varepsilon_{1}+\varepsilon_{2})/\Phi
_{0}<\varepsilon_{2}/\Phi_{2}$. These inequalities imply that a
single first soliton appears at the field
$H_{1}=4\pi^{2}\varepsilon_{1}/\varphi_{0} \Phi_{0}$. The soliton
lattice, composed of alternating solitons of two types, penetrates
into the structure at higher field
$H_{c1}=4\pi(\varepsilon_{1}+\varepsilon_{2})/\Phi_{0}$. In the
case $d_{0}=d_{\pi}$, the fields $H_{1}$ and $H_{c1}$ coincide.

An interesting possibility to have an equilibrium $\varphi_{0}$
Josephson vortex is realized at the boundary between an
alternating structure of $0$- and $\pi$- JJs and usual $0$-JJ. We
assume that the alternating structure occupies the region $x<0$,
while the $0$-JJ with the Josephson penetration length $\Lambda$
occupies a positive semi-axis $x>0$. In this case the phase
difference must vary continuously from $\varphi_{0}$ for
$x\rightarrow-\infty$ to zero for $x\rightarrow+\infty$. Due to
the condition $\lambda_{\varphi}\gg\Lambda\thicksim\lambda_{J}$,
the value $\varphi(0)\ll1$, which means that the boundary vortex
is almost completely localized in region $x<0$ and it is
equivalent to half of the first soliton (\ref{Shape1}). The
metastable $\pi -2\varphi_{0}$ boundary vortex is also possible
and in the limit $\lambda_{\varphi}\gg\Lambda$ its shape
corresponds to the second soliton (\ref{Shape2}).

Consider a structure with the total length $L$ smaller that the
screening length $\lambda_{\mathrm{eff}}$ in an external magnetic
field $H$, smaller than $\Phi_{0}/(d_{0}+d_{\pi})t$. In this case
the coarse-grained approximation for the phase is justified and we
can use the current-phase relation (\ref{Phase-Current}) where the
coarse-grained phase depends on the coordinate as
$\varphi(x)=hx+\beta$ with $h=2\pi tH/\Phi_{0}$ and $\beta$ is an
arbitrary phase shift. The total current per unit thickness
flowing through the structure is given by
\[
I(\beta)\!=\!\frac{j_{c}d_{0}^{2}}{12\lambda_{J}^{2}}\!\int\limits_{-L/2}
^{L/2}dx\left(  f\sin\left(  hx+\beta\right)  -\frac{\sin\left(
2(hx+\beta)\right)  }{2}\right).
\]
Calculating the integral, we reduce it to the form
\begin{equation}
I(\beta)=\frac{I_{0}}{\eta}\sin\left(  \beta\right)  \sin\eta\left(
f-\cos\left(  \beta\right)  \cos\eta\right)  ,
\end{equation}
where $I_{0}=j_{c}Ld_{0}^{2}/(12\lambda_{J}^{2})$ and
$\eta=hL/2=\pi\Phi/\Phi _{0}$ and $\Phi=tLH$ is the total flux
through the structure. The current reaches a maximum at
\begin{equation}
\cos\left(  \beta\right)  =\frac{\cos\varphi_{0}-\sqrt{\cos^{2}
\varphi_{0}+8\cos^{2}\left(  \eta\right)  }}{4\cos\left(
\eta\right)
}.\label{betas}
\end{equation}
Shapes of the field dependencies of the critical current at
different values of $\varphi_{0}$ are plotted in the right panel
of Fig.\ \ref{Fig-j_phi_j_h}. These dependencies differ
significantly from the Fraunhofer dependence in usual JJs. They
have large a component with half of the main period. At
$\varphi_{0}=\pi/2$ the dependence has the Fraunhofer shape but
the period is two times smaller than in usual case. Note that the
observed in the alternating $0$-$\pi $ junction structure
\cite{Smilde} finite critical current at zero magnetic field is
naturally obtained in the framework of our analysis. The
theoretical analysis in Ref.\ \onlinecite{Smilde}, giving zero
current at $H=0$, is incomplete, because it does not take into
account the current term $\propto \sin (2\varphi)$ coming from the
rapidly oscillating phase.

In conclusion, the possibility to realize the transition into the
$\varphi -$junction state with decreasing the temperature from
$T_{c}$, may be very helpful for experimental verification of the
predicted effects. In particular, the observation of the striking
non-monotonous variation of the screening length with temperature
would provide an unambiguous proof of such a transition. Also, the
studies of the fine structure of the critical current dependence
vs magnetic field and the observation of a periodicity two times
smaller than that expected for standard JJ, could be of
considerable interest. Finally, note that the SQUID microscope may
directly probe the specific shapes of the partial flux vortices.
We would like to thank A.\ Lopatin and W.\ K.\ Kwok for critical
reading of the manuscript. This work was supported by the U.\ S.\
DOE, Office of Science, under contract \# W-31-109-ENG-38.

\end{document}